\documentclass{jetpl_arXiv11096624} 
\twocolumn

\newcommand{\beq}{\begin{equation}}
\newcommand{\eeq}{\end{equation}}
\newcommand{\beqa}{\begin{eqnarray}}
\newcommand{\eeqa}{\end{eqnarray}}
\newcommand{\bsubeqs}{\begin{subequations}}
\newcommand{\esubeqs}{\end{subequations}}

\title{\vspace*{4mm}Superluminal neutrino and spontaneous breaking of Lorentz invariance}

\rtitle{Superluminal neutrino and spontaneous breaking of Lorentz invariance}

\sodtitle{Superluminal neutrino and spontaneous breaking of Lorentz
invariance}

\author{F.R. Klinkhamer $^{\#}$\/$^{1)}$
and G.E. Volovik $^{*+}$\/\thanks{\:e-mail: frans.klinkhamer@kit.edu,
volovik@boojum.hut.fi}}

\rauthor{F.R. Klinkhamer, G.E.Volovik}

\sodauthor{Klinkhamer, Volovik}

\address{$^{\#}$ Institute for Theoretical Physics, University of
Karlsruhe, Karlsruhe Institute of Technology, 76128 Karlsruhe, Germany
\\
$^{*}$ Low Temperature Laboratory, Aalto University
P.O. Box 15100, FI-00076 AALTO, Finland
\\
$^+$ L.D. Landau Institute for Theoretical Physics, Russian Academy of
Sciences, Kosygina 2, 119334 Moscow, Russia
  }

\dates{October 10, 2011}{*}


\abstract{Generally speaking, the  existence of
a superluminal neutrino can be attributed either to
re-entrant Lorentz violation at ultralow energy
from intrinsic Lorentz violation at ultrahigh energy
or to spontaneous breaking of fundamental Lorentz invariance
(possibly by the formation of a fermionic condensate).
Re-entrant Lorentz violation in the neutrino sector has been
discussed elsewhere. Here, the focus is on mechanisms
of spontaneous symmetry breaking.}

\begin{document}

\maketitle

It is possible that OPERA's claimed discovery~\cite{OPERA2011}
of a superluminal muon-type neutrino
does not come from the violation of Lorentz invariance
but from unknown factors in
the clock-synchronization process~\cite{Contaldi2011}
or from a purely statistical effect~\cite{Alicki2011}.
In fact, it has been shown~\cite{CohenGlashow2011}
that OPERA's claimed value $(v_{\nu_\mu}-c)/c\sim 10^{-5}$
is ruled out by the expected but unobserved energy losses from
electron-positron-pair emission ($\nu_\mu\to \nu_\mu+e^{-}+e^{+}$),
at least, as long as there exists a preferred frame
from the Lorentz violation.

Still, the claim by OPERA has provided new impetus for
the discussion on the possible sources of Lorentz violation.
In order to engage in this discussion, let us
assume that OPERA's result is correct qualitatively
(existence of a superluminal muon-neutrino) even if not
quantitatively (most likely, $|v_{\nu_\mu}-c|/c\ll 10^{-5}$).

Condensed-matter physics, which possesses an analog of Lorentz
invariance (LI), now suggests several different scenarios
of Lorentz violation (LV). Among them are:
\begin{enumerate}
  \item[(1a)]
LI is not a fundamental symmetry but an approximate symmetry
which emerges at low energies and is violated at ultrahigh
energies (cf.~\cite{ChadhaNielsen1982}).
  \item[(1b)]
Intrinsic LV at ultrahigh energies
gives an emergent Lorentz-invariant theory at lower energies
but ultimately, at or below an ultralow energy scale,
induces a re-entrant violation of LI
(see, e.g., Sec. 12.4 of~\cite{Volovik2003}).
  \item[(2)]
LI is fundamental but broken spontaneously
(see, e.g.,~\cite{Bjorken1963,Jenkins2004} and references therein).
\end{enumerate}

In this Letter,
we discuss the spontaneous breaking of Lorentz invariance (SBLI),
that is, the spontaneous  appearance of a preferred frame in the vacuum,
which can be derived from Lorentz-invariant physical laws.
The order parameter
of SBLI can be a vector field $b^\alpha$ (for example, the vector field of
Fermi-point splitting~\cite{KlinkhamerVolovikIJMPA2005,Volovik2007}
or an aether-type velocity field~\cite{Jacobson2008}),
an emergent tetrad-type field
$e^\alpha_a$~\cite{Akama1978,Volovik1986,Wetterich2004,Diakonov2011},
or any other field which is covariant but not invariant
under Lorentz transformations.

If SBLI occurs only in the neutrino sector,
which interacts weakly with the charged-matter sector, then SBLI
has no direct impact on this other matter
(certain indirect quantum-loop effects can be suppressed
by near-zero mixing angles).
The non-neutrino matter essentially
does not feel the existence of the preferred reference frame.
In fact, it is very well possible that
SBLI occurs only for the neutrino field, because the other
fermions have already experienced electroweak symmetry breaking phase
transitions and are too heavy for any further type of symmetry breaking.

In condensed-matter physics, the re-entrant violation of LI, as well as
Fermi-point splitting (FPS), follow from general topological properties of the
vacuum (ground state) in 3--momentum space.
Condensed matter provides many examples of homogeneous vacuum states,
which have
nontrivial topology in 3--momentum space~\cite{Volovik2003}.
Among them  is a class of vacua which have Fermi points,
exceptional points in 3--momentum space where
the energy of fermionic excitations is nullified.
Such a Fermi point (alternatively called Dirac or Weyl point)
has a topological invariant. The existence of the Fermi point
is thus protected by topology, or by the combined action of
topology and symmetry.
The Fermi point is robust to small perturbations of the system.
In turn, different Fermi points may collide,
annihilate, and split again,
but their total topological charge is conserved.
In this respect, Fermi points in 3--momentum space
behave as topologically-charged 't Hooft--Polyakov magnetic monopoles
in real space.
The splitting or recombination of Fermi points represents
a topological quantum phase transition. This type of quantum phase
transition takes place, for example, in graphene,
graphite, etc. (see, e.g., Figs.~4 and 6 in~\cite{HeikkilaVolovik2010}
for the splitting of a degenerate
Fermi point into 3 and 4 elementary points, respectively).

Figure~8 in the review~\cite{Volovik2007}
(which elaborates on the discussion of the original research
paper~\cite{KlinkhamerVolovikIJMPA2005})
illustrates the special role of neutrinos.
From the momentum-space topology of Fermi points, it follows that
the phase transition away from the symmetric vacuum of the
Standard Model with massless fermions may occur in two ways:
either coinciding Fermi points with opposite
topological charge annihilate each other, giving rise
to a Dirac mass (the process commonly known as the Higgs mechanism),
or coinciding Fermi points do not annihilate but split in momentum space,
giving rise to Lorentz violation~\cite{KlinkhamerVolovikIJMPA2005}.
As mentioned above,
it is possible that this FPS process only occurs for neutrinos,
since all other particles have
already obtained Dirac masses via the Higgs mechanism.
After this splitting, the energy spectra of the left- and right-handed
neutrino are given by the following expressions
(the small neutrino mass can be neglected for the conditions
relevant to the OPERA experiment):%
\begin{subequations}\label{Spectrum-FPS}
\begin{eqnarray}
g^{\alpha\beta}\,\Big(c\,p_\alpha - \widetilde{b}_\alpha^{\rm \,L}\Big)\,
             \Big(c\,p_\beta - \widetilde{b}_\beta^{\rm \,L}\Big) &=&0\,,
\\[2mm]
g^{\alpha\beta}\,\Big(c\,p_\alpha - \widetilde{b}_\alpha^{\rm \,R}\Big)\,
            \Big(c\,p_\beta - \widetilde{b}_\beta^{\rm \,R}\Big)&=&0\,,
\end{eqnarray}
\end{subequations}
with $c$ the velocity of light \textit{in vacuo}.
In \eqref{Spectrum-FPS}, we have put a tilde on the dimensional
vector $\widetilde{b}_\alpha$ in order to distinguish it from
the  dimensionless vector $b_\alpha$ appearing below and we allow for
$\widetilde{b}_\alpha^{\rm \,L} \neq  \widetilde{b}_\alpha^{\rm \,R}$.

The possible role of FPS for the (qualitative) OPERA
result has already been discussed in~\cite{Klinkhamer2011}.
Here, we consider two scenarios for the
spontaneous formation of a preferred reference frame.
The first scenario corresponds to the appearance of a dimensionless
4--vector $(b^\alpha)=(b^0,{\bf b})$ in the neutrino vacuum.
This 4--vector $b^\alpha$ interacts with the neutrino Dirac field
in a way
which does not violate the fundamental laws of special relativity:
\begin{eqnarray}\label{bInteraction}
S&=&\int\, d^3x \,c\,dt ~
\overline{\psi}\: b^\alpha \,(-i\,\nabla_\alpha)\:\psi\,.
\end{eqnarray}
This action term corresponds to a momentum-dependent
mass term,\footnote{At this stage, it is clear that the same
procedure can be followed with a Majorana mass term
$\overline{\psi}_L^\text{\;c}\,m\,\psi_L$ in the
action density, simply replacing the Majorana mass
$m$  by $-i\,b^\alpha\,\nabla_\alpha/c\,$.}
$M=p_\alpha\, b^\alpha/c$,
which modifies the spectrum of the neutrino as follows:
\begin{equation}
p_\alpha \,p^\alpha \equiv
E^2-c^2\,|{\bf p}|^2=\big(c\,p_\alpha\, b^\alpha\big)^2\,.
\label{Spectrum}
\end{equation}
Let us, for example, take $b^\alpha$ to be a timelike vector,
having $(b^0)^2-|{\bf b}|^2>0$ ,
and consider the particular reference frame
with ${\bf b}=0$. Assume $|b^0| < 1$.
Then, the neutrino energy spectrum becomes
\begin{equation}
E^2\,\big[1-(b^0)^2\big]=c^2\,|{\bf p}|^2\,,
\label{SpectrumTimeLike}
\end{equation}
which is superluminal for $b^0 \ne 0$.
The same happens for a spacelike vector $b^\alpha$.
In the reference frame with $b^0=0$,
the neutrino energy spectrum is given by
\begin{equation}
E^2=c^2\,|{\bf p}|^2 + c^2\,({\bf b}\cdot {\bf p})^2\,,
\label{SpectrumSpaceLike}
\end{equation}
which is both anisotropic and superluminal for ${\bf b}\ne 0$.

The 4--vector field $b^\alpha$ may emerge as the order
parameter of the neutrino condensate,
\begin{eqnarray}\label{Condensate-without-gamma5}
b^\alpha &\propto& g^{\alpha\beta}
\left< \overline{\psi}\, (-i\,\nabla_\beta)\, \psi  \right>\,,
\end{eqnarray}
in a theory with 4--fermion or multi-fermion interactions
of the following type:
\begin{subequations}\label{eq:multi-fermion-interaction}
\begin{eqnarray}
S_{\rm int} &=& \int\, d^3x \,c\,dt ~f(X)~~,
\\[2mm]
X&=& - g^{\alpha\beta}\, \big(\overline{\psi}\,\nabla_\alpha\, \psi\big) \;
       \big(\overline{\psi}\, \nabla_\beta\, \psi\big) \,,
\end{eqnarray}
\end{subequations}
with appropriate dimensional
constants entering the function $f$.
This scenario gives a possible realization of the phenomenological
Coleman--Glashow model~\cite{ColemanGlashow1997-1999}
in terms of fermionic condensates~\cite{Bjorken1963,Jenkins2004}.

The second scenario involves another type of neutrino condensate,
which also leads to SBLI.
Specifically, this neutrino condensate gives rise to a tetrad-like
field~\cite{Akama1978,Volovik1986,Wetterich2004,Diakonov2011}:
\begin{equation}
e^a_\alpha\propto
\left< \overline{\psi}\, \gamma^a (-i\,\nabla_\alpha)\, \psi  \right>\,.
\label{CondensateTetrad}
\end{equation}
The induced tetrad field $e^a_\alpha$
must be added to the original
fundamental tetrad $E^{(0)\alpha}_a={\rm diag}(-1,\,c,\,c,\,c)$,
and the fermionic action becomes
\begin{subequations}\label{ModifiedTetrad}
\begin{eqnarray}
S&=&\int\, d^3x \,c\,dt ~
E^\alpha_a\,\overline{\psi}\,\gamma^a (-i\,\nabla_\alpha)\, \psi~~,
\\[2mm]
E^\alpha_a &=& E^{(0)\alpha}_a+e_a^\alpha\,.
\end{eqnarray}
\end{subequations}
Using the induced tetrad
field $(e^\alpha_a)={\rm diag}(b_0,\,0,\,0,\,0)$ as an example,
one obtains, for
$0<b_0<1$, the superluminal neutrino velocity $v_\nu=c/(1-b_0)$.

The tetrad-type neutrino condensate \eqref{CondensateTetrad}
may provide a realization of the hypothetical spin-2 field
discussed in~\cite{DvaliVikman2011}.
A recent paper~\cite{Kehagias2011} presents another model,
where a scalar-field composite
plays a similar role as our condensate $e_a^\alpha$.
Also related may be a geometric model~\cite{PfeiferWohlfarth2011},
based on a particular class of Finsler-spacetime backgrounds,
which essentially modifies the effective metric entering
particle dispersion relations.

This completes our discussion of two possible scenarios
of spontaneous symmetry breaking to explain a superluminal
neutrino. Spontaneous breaking of Lorentz invariance
in the neutrino sector corresponds to
the appearance of a preferred frame for the relevant neutrino:
LI is violated if the neutrino momentum $p_\alpha$ is transformed but not
the vacuum field $b^\alpha$ (or $e^a_\alpha$) which is kept
at a fixed value.
Still, LI remains an exact symmetry of the physical laws:
the invariance holds if both excitations and vacuum are transformed,
that is, if the Lorentz transformation acts simultaneously on
$p_\alpha$ and $b^\alpha$  (or $e^a_\alpha$).

In these SBLI scenarios,
as well as in the FPS scenario~\cite{Klinkhamer2011},
the vacuum remains homogeneous, which is the reason
why conservation of energy and momentum is exact.
But the energy spectrum of the neutrino is modified,
which must certainly have consequences for
reactions involving neutrinos. Hence,
there must be experimental constraints on  $b^\alpha$ or  $e^a_\alpha$.
Alternatively, the study of neutrino-interaction processes may provide
valuable information on mechanisms proposed
to explain nonstandard (e.g., superluminal)
propagation properties of the neutrinos.

The advantage of the spontaneous-symmetry-breaking scenario
is that it stays fully within the realm of standard physics,
which obeys special relativity.
The multi-fermion interaction \eqref{eq:multi-fermion-interaction}
can, in principle, originate from trans-Planckian physics,
but we now have bounds~\cite{GagnonMoore2004,BernadotteKlinkhamer2007}
indicating that Lorentz invariance holds far above
the Planck energy scale, i.e., $E_{\rm LV}\gg E_{\rm Planck}$.
This suggests that, if a neutrino has superluminal motion,
it can be attributed either
to re-entrant Lorentz violation at ultralow energy due to
intrinsic (built-in) Lorentz violation at
ultrahigh trans-Planckian energies
[presumably with a re-entrance energy
of order $(E_{\rm Planck}/E_{\rm LV})^n\,E_{\rm Planck}$
for $n\geq 1$] or
to spontaneous breaking of fundamental Lorentz invariance
[possibly by the formation of a fermionic condensate].

\vfill


\begin{thebibliography}{99}

\bibitem{OPERA2011}
T. Adam {\it et al.} [OPERA Collaboration],
``Measurement of the neutrino velocity with the OPERA detector
in the CNGS beam,''
arXiv:1109.4897v1.

\bibitem{Contaldi2011}
C.R. Contaldi,
``The OPERA neutrino velocity result and the synchronisation of clocks,''
arXiv:1109.6160.

\bibitem{Alicki2011}
R. Alicki,
``A possible statistical mechanism of anomalous neutrino velocity
in OPERA experiment?,''
arXiv:1109.5727.

\bibitem{CohenGlashow2011}
A.G.~Cohen and S.L.~Glashow,
``New constraints on neutrino velocities,''
arXiv:1109.6562. 

\bibitem{ChadhaNielsen1982}
S.~Chadha and H.B.~Nielsen,
  ``Lorentz invariance as a low-energy phenomenon,''
Nucl. Phys. B {\bf 217}, 125 (1983).

\bibitem{Volovik2003}
G.E. Volovik,
{\it The Universe in a Helium Droplet},
Clarendon Press,  Oxford (2003).

\bibitem{Bjorken1963}
J.D.~Bjorken,
``A dynamical origin for the electromagnetic field,''
Annals Phys.\  {\bf 24}, 174 (1963).

\bibitem{Jenkins2004}
A. Jenkins,
``Spontaneous breaking of Lorentz invariance,''
Phys. Rev. D {\bf 69}, 105007 (2004),
arXiv:hep-th/0311127.

\bibitem{KlinkhamerVolovikIJMPA2005}
F.R. Klinkhamer and G.E. Volovik,
``Emergent CPT violation from the splitting of Fermi points,''
Int. J. Mod. Phys. A {\bf 20}, 2795 (2005),
arXiv:hep-th/0403037.

\bibitem{Volovik2007}
G.E. Volovik,
``Quantum phase transitions from topology in momentum space,''
Lect. Notes Phys.  {\bf 718}, 31 (2007),
arXiv:cond-mat/0601372.

\bibitem{Jacobson2008}
T. Jacobson,
``Einstein--aether gravity: A status report,''
PoS {\bf QG-PH}, 020 (2007),
arXiv:0801.1547.

\bibitem{Akama1978}
K. Akama,
``An attempt at pregeometry -- gravity with composite metric,''
Prog. Theor. Phys. {\bf 60}, 1900 (1978).

\bibitem{Volovik1986}
G.E. Volovik,
``Superfluid $^3$He--B and gravity,''
Physica B  {\bf 162}, 222 (1990).

\newpage
\bibitem{Wetterich2004}
C. Wetterich,
``Gravity from spinors,''
Phys. Rev. D {\bf 70}, 105004 (2004),
arXiv:hep-th/0307145.

\bibitem{Diakonov2011}
D. Diakonov,
``Towards lattice-regularized quantum gravity,''
arXiv:1109.0091.

\bibitem{HeikkilaVolovik2010}
T.T. Heikkil\"a and G.E. Volovik,
``Fermions with cubic and quartic spectrum,''
JETP Lett. {\bf 92}, 681 (2010),
arXiv:1010.0393.

\bibitem{Klinkhamer2011}
F.R. Klinkhamer,
``Superluminal muon-neutrino velocity
from a Fermi-point-splitting model of Lorentz violation,''
arXiv:1109.5671.

\bibitem{ColemanGlashow1997-1999}
S. R. Coleman  and  S. L. Glashow,
``Cosmic ray and neutrino tests of special relativity,''
Phys. Lett. B {\bf 405}, 249  (1997),
arXiv:hep-ph/9703240;
S. R. Coleman  and  S. L. Glashow,
``High-energy tests of Lorentz invariance,''
Phys. Rev. D {\bf 59}, 116008 (1999),
arXiv:hep-ph/9812418.

\bibitem{DvaliVikman2011}
G. Dvali and A. Vikman,
``Price for environmental neutrino-superluminality,''
arXiv:1109.5685.

\bibitem{Kehagias2011}
A.~Kehagias,
``Relativistic superluminal neutrinos,''
arXiv:1109.6312.

\bibitem{PfeiferWohlfarth2011}
C. Pfeifer and M.N.R. Wohlfarth,
``Beyond the speed of light on Finsler spacetimes,''
arXiv:1109.6005.  

\bibitem{GagnonMoore2004}
O. Gagnon and G.D. Moore,
``Limits on Lorentz violation from the highest energy cosmic rays,''
Phys. Rev. D {\bf 70},  065002 (2004),
arXiv:hep-ph/0404196.

\bibitem{BernadotteKlinkhamer2007}
S. Bernadotte and F.R. Klinkhamer,
``Bounds on length scales of classical spacetime foam models,''
Phys.\ Rev.\  D {\bf 75}, 024028 (2007), arXiv:hep-ph/0610216.


\end{thebibliography}
\end{document}